\documentclass[twocolumn,showpacs,a4paper,superscriptaddress,nofootinbib,tightenlines,floats]{revtex4}
\usepackage{bm}
\usepackage{latexsym}
\usepackage{dcolumn}
\usepackage{amsfonts,amssymb}
\usepackage{graphicx,epsfig}
\usepackage{psfrag}
\usepackage{amsmath,amssymb}

\begin{document}

\title{\textbf{Scalar Field Dark Energy Perturbations and their Scale
    Dependence}}

\author{Sanil Unnikrishnan}
\email{sanil@physics.du.ac.in}
\affiliation{Department of Physics \& Astrophysics, University of Delhi,
 Delhi : 110007, India}

\author{H. K. Jassal}
\email{hkj@mri.ernet.in}
\affiliation{Harish-Chandra Research Institute, Chhatnag Road,
Jhunsi, Allahabad 211 019, India}

\author{T. R. Seshadri}
\email{trs@physics.du.ac.in}
\affiliation{Department of Physics \& Astrophysics, University of Delhi,
 Delhi : 110007, India}


\date{\today}


\begin{abstract}

We estimate the amplitude of perturbation in dark energy at different
length scales for a quintessence model with an exponential potential.
It is shown that on length scales much smaller than  hubble radius,
perturbation in dark energy is negligible in comparison to that in
in dark matter.
However, on scales comparable to the hubble radius
($\lambda_{p}>1000\mathrm{Mpc}$) the perturbation in dark energy in general
cannot be neglected.
As compared to the $\Lambda$CDM model,  large scale matter power spectrum is
suppressed in a generic quintessence dark energy model.
We show that on  scales $\lambda_{p} < 1000\mathrm{Mpc}$, this suppression is
primarily due to different background evolution compared to $\Lambda$CDM
model.
However, on much larger scales perturbation in dark energy can effect
matter power spectrum significantly.
Hence this analysis can act as a discriminator between $\Lambda$CDM model and
 other generic dark energy models with $w_{de} \neq -1$.

\end{abstract}

\maketitle

\section{Introduction}
Cosmological observations  suggest that about $70\%$ content of our
universe is made of a form of matter which drives the accelerated expansion of
the universe\cite{obs_proof}.
These observations include Supernova type Ia observations
\cite{nova_data1}, observations of Cosmic Microwave Background
\cite{boomerang,wmap_params,2003Sci...299.1532B} and
large scale structure \cite{2df, sdss, 2004PhRvD..69j3501T}.
The accelerated expansion of the universe can  of course be explained by
introducing a cosmological constant $\Lambda$ in the Einstein's equation
\cite{ccprob_wein,review3}.
However, the cosmological constant model is plagued by the fine turning
problem \cite{ccprob_wein}.
This has motivated the study of dark energy models to explain the current
accelerated expansion of the universe.
The simplest model as an alternative to cosmological constant model is to
assume that this accelerated expansion is driven by a canonical scalar field
with a potential $V(\phi)$.
This class of dark energy models are known as quintessence models and the
scalar field is known as a quintessence field.
Various quintessence models have been studied in literature
\cite{quint1,ferreira,liddle}.
There exists another class of string theory inspired scalar field dark energy
models known as tachyon models \cite{tachyon1,2003PhRvD..67f3504B}.
Models of dark energy which allow $w<-1$ are known as phantom
models \cite{2002PhLB..545...23C}.
Phantom type dark energy can also be realized in a scalar tensor theory of
gravitation. (See for example Ref.\cite{STG}.)
Other scalar field models include k-essence field \cite{2001PhRvD..63j3510A},
branes \cite{brane1}, Chaplygin gas model and its generalizations
\cite{chaply}.
There are also some phenomenological models \cite{water}, field theoretical
and reorganization group based models  (see e.g. \cite{tp173}), models that
unify dark matter and dark energy \cite{unified_dedm1}, holographic dark
energy models \cite{HGDE}  and many others like those based on horizon
thermodynamics (e.g. see \cite{2005astro.ph..5133S}).
For reviews of dark energy models see for instance Ref.\cite{DEreview} and for
constraining parameters using observations see Ref.\cite{param_fit}.

Homogeneous dark energy distribution leads to accelerated expansion of the
Universe which, in turn, governs the luminosity distance and angular diameter
distance.
The rate of expansion also
influences growth of density perturbations in the universe.
This is evident from  the abundance of rich clusters of galaxies and their
evolution and the Integrated Sachs Wolfe effect \cite{isw0}.

In this paper we present a set of argument which lead to the conclusion
that inhomogeneous dark matter with homogeneous dark energy at all length
scales is inconsistent with Einstein's equation if $p_{de} \neq -\rho_{de}$.
We further analyze how dark matter power spectrum is influenced by perturbation
 in dark energy with an evolving equation of state parameter.

Dark energy perturbations have been extensively studied in the linear
approximation \cite{weller_lewis,bean_dore,depert}.
It was shown in Ref.\cite{weller_lewis} that dark energy perturbations affect
the low $l$ quadrapole in the CMB angular power spectrum through the ISW
effect.
This analysis was done for a constant equation of state parameter.
For models with $w>-1$ this effect is enhanced while for phantom like models
it is suppressed.
In these models dark matter perturbations and dark energy perturbations are
anti-correlated for large effective sound speeds.
This anti-correlation is a gauge dependent effect
\cite{bean_dore}.
Detailed studies in dark energy perturbations also include
\cite{chpgas_pert,sph_coll}.
Clustering of dark energy within matter over density and voids were studied by
Mota et al\cite{mota}.

In this paper we study the evolution of perturbation in dark energy in a quintessence model which results
in an evolving equation of state parameter different from that considered in Refs.\cite{amendola1, weller_lewis}.
We use a specific model of scalar field dark energy with an
exponential potential.
We find that although for scales much smaller than hubble radius the
perturbation in dark energy is small, for scales $\geq H^{-1}$ the
perturbation in dark energy can be comparable to that in matter.
Hence, although on small scales ($<< H^{-1}$) the dark energy can be treated
as homogeneous, one has to take into account the perturbation in dark energy
over scales $\sim H^{-1}$ if $w_{de} \neq -1$.
Clearly, in the specific case of $w_{de} = -1$ the dark energy is homogeneous
at all scales.

We choose to work in the longitudinal gauge as in that case we can directly
relate the metric perturbation $\Phi$ to the gravitational potential
perturbation.
For a specific model, we investigate how quintessence  dark energy influences
matter power spectrum.
We show that on scales $\lambda_{p} < 1000\mathrm{Mpc}$, the matter power
spectrum is not significantly affected whether or not  we include
fluctuations in the quintessence field in  perturbation equations.
However, on much larger scales, including or excluding fluctuations in
quintessence field does result in significant changes in the matter power
spectrum.

This paper is organized as follows.
In Section \ref{sec::inhomogeneous DE} we  discuss the background
cosmology for matter and the scalar field system and describe the
cosmological perturbation equation in longitudinal gauge for this system.
In  Section \ref{sec::Numerical Solutions} we obtain the
numerical solution of the perturbation equation to determine the ratio of the
perturbations in dark energy to the perturbations in matter at various length
scales.
Section \ref{sec::Conclusions} summarizes the results.


\section{Inhomogeneous matter and Dark energy}\label{sec::inhomogeneous DE}
We shall consider a system of minimally coupled matter (dark + baryonic) and
canonical scalar field \footnote{We shall denote scalar field by $\phi$ and
the metric perturbation in the longitudinal gauge by $\Phi$} with Lagrangian
of the form :
\begin{equation}
\mathcal{L}_{\phi} =\frac{1}{2}\partial_\mu\phi\partial^\mu\phi -
V(\phi)\label{eqn::lagrangian scalar field}
\end{equation}

As a simple example, we consider a scalar field potential of the form
\cite{ratra88,burd,copeland,exp_pot}:
\begin{equation}
V(\phi) = V_{o}\exp\big[-\sqrt{\lambda}
  \frac{\phi}{M_{p}}\big]\label{eqn::potential Vphi}
\end{equation}
Here $\lambda$ and $V_{o}$ are two  parameters of the potential and $M_{p} =
(8 \pi G)^{-1/2}$ is the Planck mass.
This potential leads to scaling solutions\cite{ferreira,liddle,copeland}.
However, for treating exponential potential as a possible candidate for dark
energy, we require that the scalar field should not enter the scaling
regime.
This is possible to achieve  by restricting the choice of the parameter in the
exponential potential \cite{copeland}.

A spatially flat homogeneous and isotropic line element is described by the
FRW metric of the form :
\begin{equation}
ds^{2} = dt^{2} - a^{2}(t)[dx^{2} + dy^{2} + dz^{2}]\label{line element}
\end{equation}

For the system of pressureless matter and  quintessence with exponential
potential the background evolution $a(t)$  is completely determined by solving
the following Friedmann equation  and the Klein Gordon equation :
\begin{equation}
\frac{\dot{a}^2}{a^2} = \frac{8 \pi G }{3}\Big[\rho_{m}a^{-3} +
  \frac{1}{2}\dot{\phi}^{2} + V_{o}\exp(-\sqrt{\lambda}
  \frac{\phi}{M_{p}})\Big]\label{eqn::background eqn 1}
\end{equation}
and
\begin{equation}
\ddot{\phi} + 3\frac{\dot{a}}{a}\dot{\phi} -
\frac{\sqrt{\lambda}}{M_{p}}V_{o}\exp\big[-\sqrt{\lambda}
  \frac{\phi}{M_{p}}\big] = 0 \label{eqn::background eqn 2}
\end{equation}

\subsection{Equations for Perturbations}
For analyzing perturbations in scalar fields and matter we work in the
longitudinal gauge \cite{bardeen PRD 1980, Kodama, mukhanov 1992} (For a recent
pedagogical review see, \cite{tp_rev}).
 For scalar field and for pressureless matter
(described as perfect fluid), the perturbed energy momentum tensor has no
anisotropic term \textit{i.e.} $\delta T^{i}_{\hspace{0.1cm}j}$ is diagonal
(where $ i , j = 1,2 ,3$)\footnote{Please note that through out this paper
  $\mu,\nu = 0,1,2,3$ and $ i, j = 1, 2 ,3$}.
Einstein's equations would then imply that the scalar metric perturbations
in this gauge are completely described by a single scalar variable $\Phi$.
For such a system, perturbed FRW metric in longitudinal gauge attains the form
\cite{mukhanov 1992}:
\begin{equation}
    ds^{2} = ( 1 + 2\Phi)dt^{2} - a^{2}(t)(1 - 2 \Phi)[dx^2 + dy^2 +
    dz^2]\label{eqn::longitudinal gauge}
\end{equation}

It is a good approximation to treat Dark Matter, baryonic matter, etc. as
perfect fluids.
The energy momentum tensor of a perfect fluid is described as:
\begin{equation}
T^{\mu}_{\hspace{0.2cm}\nu} = (\rho + p)u^{\mu}u_{\nu} -
p\delta^{\mu}_{\hspace{0.2cm}\nu} \label{eqn::Tpf}
\end{equation}

Perturbations in the energy density $\rho$, pressure $p$ and the four velocity
$u^{\mu}$ are defined as:
\begin{eqnarray}
\rho(t,\vec{x}) &=&  \rho_{o}(t) + \delta \rho(t,\vec{x}) \label{eqn::pfa}\\
p(t,\vec{x}) &=&  p_{o}(t) + \delta p(t,\vec{x}) \label{eqn::pfb}\\
u^{\mu} &=& \hspace{0.1cm}^{(o)} u^{\mu} + \delta u^{\mu} \label{eqn::pfc}
\end{eqnarray}
where  $\hspace{0.1cm}^{(o)} u^{\mu} = \{1,0,0,0\}$;  $\rho_{o}(t)$ and
$p_{o}(t)$ are average values of the energy density and pressure
respectively on a constant time hypersurface.
The peculiar velocity is given by  $\delta u^{i}$.
Substituting Eqs. (\ref{eqn::pfa}), (\ref{eqn::pfb}) and (\ref{eqn::pfc}) in
Eq. (\ref{eqn::Tpf}) and neglecting second and higher order perturbation terms
we  get:
\begin{eqnarray}
\delta T^{0}_{\hspace{0.2cm}0}  &=&  \delta \rho
\label{eqn::perturbed density fluid}\\
\delta T^{i}_{\hspace{0.1cm}0}  &=&  \big (\rho_{o} + p_{o}\big )\delta
  u^{i}\label{eqn::peculier velosity fluid} \\
\delta T^{i}_{\hspace{0.1cm}j}  &=&  -\delta p
  \delta^{i}_{\hspace{0.1cm}j}\label{eqn::perturbed pressure fluid}
\end{eqnarray}
For scalar (quintessence) field with Lagrangian of the form
Eq.(\ref{eqn::lagrangian scalar field}), we define the perturbations as :
\begin{equation}
\phi(\vec{x},t) = \phi_{o}(t) + \delta\phi(\vec{x},t)
\label{eqn::scalar field perturbation}
\end{equation}
where $\phi_{o}(t)$ is the average value of the scalar field on the constant
time hypersurface.
The energy momentum tensor for the scaler field is given by
\begin{eqnarray}
T^{\mu}_{\hspace{0.2cm}\nu} = \partial^{\mu}\phi\partial_{\nu}\phi - \mathcal{L}_{\phi}\delta^{\mu}_{\hspace{0.2cm}\nu}\label{eqn::EM tensor scalar field}
\end{eqnarray}
Substituting Eq.(\ref{eqn::scalar field perturbation}) in Eq.(\ref{eqn::EM tensor scalar field}) and subtracting the homogeneous part in the energy momentum tensor we get
\begin{eqnarray}
\delta T^{0}_{\hspace{0.2cm}0} &=& \delta\rho_{\phi} =
\dot{\phi}_{o}\dot{\delta\phi} - \Phi\dot{\phi}_{o}^2  +
V'(\phi_{o})\delta\phi\label{eqn:: perturbed density scalar field}\\
\delta T^{i}_{\hspace{0.01cm}j} &=& -\delta
p_{\phi}\delta^{i}_{\hspace{0.01cm}j} = -\left[\dot{\phi}_{o}\dot{\delta\phi} -
  \Phi\dot{\phi}_{o}^2 - V'\delta\phi \right]\delta^{i}_{\hspace{0.01cm}j}
\label{eqn:: perturbed pressure scalar field}\\
\delta T^{0}_{\hspace{0.2cm}i}  &=& (\rho_{_{\phi_o}} + p_{_{\phi_o}})\delta
u_{i_{(\phi)}} = \dot{\phi}_{o}\delta \phi_{,i}
\label{eqn::peculiar velocity scalar field}
\end{eqnarray}

\subsection{Inhomogeneous Dark energy}
The existence of inhomogeneity  in non relativistic matter (dark + baryonic)
is evident from direct observations.
We ask : is it reasonable to assume that
dark energy is in general homogeneous given the observational fact that matter (dark +
baryonic) is clustered?
Dark energy has equation of state $p < -\frac{1}{3}\rho$, leading to the
accelerated expansion of the universe.
The fluid with such an equation of state
 behaves gravitationally as a repulsive form of matter and opposes gravitational
clustering.

Our assumption that matter and scalar field are minimally coupled implies that
energy momentum tensor for both matter and scalar
field are individually conserved.
This would then imply that for both these components the variables defining
the perturbations $\delta \rho$, $\delta p$ and $\delta u^{i}$ would satisfy
the following two equations\footnote{Note that for   scalar field $\delta
T^{0}_{\hspace{0.2cm}i}  = \dot{\phi}_{o}\delta \phi_{, i}$.
This could also be written as $\delta T^{0}_{\hspace{0.2cm}i}  =  \big
(\rho_{\phi_{o}} + p_{\phi_{o}}\big )\delta u_{i_{(\phi)}}$ where   $\delta
u_{i_{(\phi)}} = \dot{\phi}^{-1}\delta \phi_{, i}$ .} :

\begin{equation}
\dot{\delta \rho} + (\rho_{o}  + p_{o})\vec{\nabla} .\vec{\delta u} +
3\frac{\dot{a}}{a}(\delta \rho + \delta p) - 3\dot{\Phi}(\rho_{o} + p_{o}) = 0
\label{eqn::perturned energy conservation equation}
\end{equation}
\begin{equation}
(\rho_{o} + p_{o})\dot{\delta u}_{i} + \dot{p}_{o}\delta u_{i} - \delta p_{,i}
  - \Phi_{,i}(\rho_{o} + p_{o}) = 0 \label{eqn::euler equation}
\end{equation}

If dark energy with $w \neq -1$ were to be homogeneously distributed,
\textit{i.e} if $\delta
T^{\mu}_{\hspace{0.2cm}\nu (DE)} = 0$, then,
Eqs.(\ref{eqn::perturned energy conservation equation}) and
(\ref{eqn::euler equation}) imply that the gravitational potential does not
depend on space and time (\textit{i.e.} $\Phi = $ constant).
If this is the case then in the line element (\ref{eqn::longitudinal gauge}),
we can  rescale time and space coordinate such that Eq.(\ref{eqn::longitudinal
  gauge}) becomes FRW metric (\ref{line element}).
This would then imply that $\delta \rho_{m} = 0$.
Hence homogeneously
distributed dark energy at all length scales would naturally imply that matter
is also distributed homogeneously.
As we see structures over different scales,
this is an observational evidence that the matter in the universe
is clearly not homogeneously distributed.
Hence, if we assume that dark energy with $w \neq -1$
is homogeneously distributed at all length scales,
then it is inconsistent with the observed features of the Universe\cite{weller_lewis, bean_dore, depert}.

\begin{figure}[t]
\begin{center}
\psfig{file = 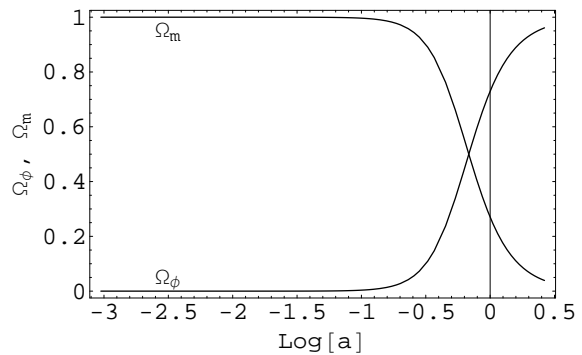, width = 3in}
\caption{This plot shows the variation of $\Omega_{m}$ and $\Omega_{\phi}$ as
 a function of scale factor.  We have chosen $\lambda = 1$.
 In the x-axis, ``Log" refers to the logarithm to base 10.}
\label{plot::omega}
 \end{center}
\end{figure}

\subsection{Linearized Einstein's Equation}
The perturbed Einstein's Equation about a flat FRW metric are given by
$\delta G^{\mu}_{\hspace{0.2cm}\nu} = 8 \pi G \hspace{0.1cm}\delta
T^{\mu}_{\hspace{0.2cm}\nu}$. In our case $ \delta T^{\mu}_{\hspace{0.2cm}\nu}
= \delta T^{\mu}_{\hspace{0.2cm}\nu (\mathrm{matter})} + \delta
T^{\mu}_{\hspace{0.2cm}\nu (\phi)}$.
Since matter has negligible pressure we set $\delta
T^{i}_{\hspace{0.1cm}j(\mathrm{matter})} = 0$.
Fluctuation in pressure is contributed only by the scalar field.
Calculating the perturbed Einstein's tensor $\delta
G^{\mu}_{\hspace{0.2cm}\nu}$ from the line element (\ref{eqn::longitudinal
gauge}) and substituting $\delta T^{\mu}_{\hspace{0.2cm}\nu}$ from
Eqs.(\ref{eqn::perturbed density fluid}) to
(\ref{eqn::perturbed pressure fluid}) and from  Eqs.(\ref{eqn:: perturbed
  density scalar field}) to (\ref{eqn:: perturbed pressure scalar field}), we
obtain the following linearized Einstein's equations :

\begin{eqnarray}
3\frac{\dot{a^2}}{a^2} \Phi + 3 \frac{\dot{a}}{a}\dot{\Phi}+
\frac{k^2\Phi}{a^2} &=& -4\pi G\left[ \delta \rho_{m} +\dot{\phi}_{o}\dot{\delta\phi}- \right.\nonumber\\
&&\left.  \Phi\dot{\phi}_{o}^2  +
  V'(\phi_{o})\delta{\phi} \right] \label{eqn::linerized einstein eqn 1}\\
\ddot{\Phi} + 4\frac{\dot{a}}{a}\dot{\Phi} + \big (2\frac{\ddot{a}}{a} +
\frac{\dot{a}^2}{a^2}\big )\Phi &=& 4\pi G\left[\dot{\phi}_{o}\dot{\delta\phi}
  \right.  \nonumber\\
&& \left. -\Phi\dot{\phi}_{o}^2   - V'(\phi_{o})\delta{\phi}\right]
\label{eqn::linerized einstein eqn 3}\\
\dot{\Phi} + \frac{\dot{a}}{a}\Phi &=& 4 \pi G \big(\rho_{o}a^{-3}v_{m} +
\dot{\phi}_{o}\delta \phi\big)\label{eqn::linerized einstein eqn 2}
\end{eqnarray}
where $V'(\phi_{o}) = \partial V(\phi_{o})/\partial\phi_{o}$ and $v_{m}$ is
the potential for the matter peculiar velocity such that $\delta u_{i} =
\nabla_{i}v_{m}$.
Since these equations are linear, we have fourier decomposed the perturbed
quantities such as $\Phi$, $\delta \phi$,  $\delta \rho_{m}$ and $v_{m}$ and
replaced $\nabla^{2}$ by $-k^{2}$, where $k$ is the wave number defined as $k
= 2\pi/\lambda_{p}$ and $\lambda_{p}$ is the comoving length scale of
perturbation.
In these equations all the perturbed quantities  correspond to the amplitude
of perturbations in the $k^{th}$ mode.

\begin{figure}
\begin{center}
\psfig{file = 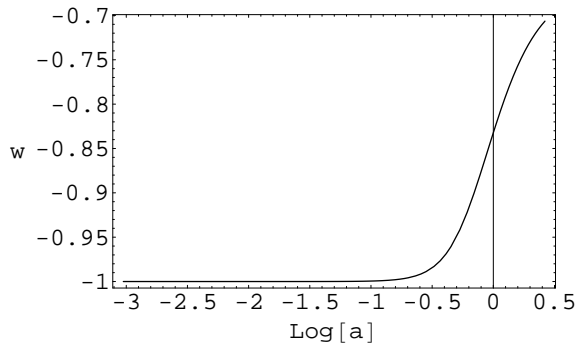, width = 3in}
\caption{This plot shows the variation of equation of state parameter
  $w_{\phi}$ as a function of scale factor. We have chosen $\lambda = 1$.
  In the x-axis, ``Log" refers to the logarithm to base 10.}
\label{plot::w}
\end{center}
\end{figure}

Eq.(\ref{eqn::linerized einstein eqn 3}) is the dynamical equation for the
metric perturbation $\Phi$ and the perturbation in the scalar field turns out
to be the driving term.
The unknown variables are $\Phi$, $\delta \phi$ and $\delta \rho_{m}$.
Once $\Phi (t)$ and  $\delta \phi (t)$ are known then the potential for matter
peculiar velocity $v_{m}$ can be determined using
Eq.(\ref{eqn::linerized einstein eqn 2}).
Also it is interesting to note that once $\Phi (t)$ and  $\delta \phi (t)$ are
known then even the matter density perturbation $\delta \rho_{m}$ can be
determined using Eq. (\ref{eqn::linerized einstein eqn 1}).
Hence for such a system of pressureless matter and scalar field, the dynamics
of perturbations is uniquely determined if we know the solution $\Phi (t)$ and
$\delta \phi (t)$.
Hence we need just two second order equations connecting $\Phi(t)$, $\delta
\phi(t)$.
We choose Eq. (\ref{eqn::linerized einstein eqn 3}) as one of these equations.
In addition to this, the dynamical equation for the perturbations in the
scalar field $\delta \phi (t)$ is obtained from the scalar field lagrangian (\ref{eqn::lagrangian scalar field}) and this is given by :
\begin{equation}
\ddot{\delta \phi} + 3\frac{\dot{a}}{a}\dot{\delta \phi} + \frac{k^2 \delta
  \phi}{a^2} + 2\Phi V'(\phi_{o})  -4\dot{\Phi}\dot{\phi_{o}} +
V''(\phi_{o})\delta \phi = 0 \label{eqn:: perturbed klein gordon eqn}
\end{equation}

For any quintessence potential $V(\phi)$, in a system of matter and
scalar field, one can solve the coupled Eqs.(\ref{eqn::linerized einstein eqn
  3}) and (\ref{eqn:: perturbed klein gordon eqn}) to study the behavior of
the perturbed system.
Once the solution  $\Phi (t)$ and  $\delta \phi (t)$ is obtained, we can then
calculate the fractional density perturbation defined as:
\begin{equation}
\delta = \frac{\delta \rho}{\rho_{o}},
\end{equation}
for both matter as well as scalar field from Eqs.
(\ref{eqn:: perturbed density scalar field}) and
(\ref{eqn::linerized einstein eqn 1}).
This is given by :
\begin{eqnarray}
\delta_{\phi} &=& \frac{1}{ \frac{1}{2}\dot{\phi}_{o}^2 +
  V(\phi_{o})}\Big[\dot{\phi}_{o}\dot{\delta\phi} - \Phi\dot{\phi}_{o}^2 +
  V'(\phi_{o})\delta{\phi}\Big]\label{eqn::delta varphi}\\
\delta_{m} &=& -\frac{1}{4 \pi G \rho_{m_{o}}a^{-3}}\Big
  \{3\frac{\dot{a^2}}{a^2} \Phi +3 \frac{\dot{a}}{a}\dot{\Phi} +
  \frac{k^2\Phi}{a^2} \Big \}\nonumber \\
&& + \frac{\delta_{\phi}}{\rho_{m_{o}}a^{-3}}\Big[ \frac{1}{2}\dot{\phi}_{o}^2
  + V(\phi_{o})\Big]\label{eqn::deltaM}
\end{eqnarray}

Using these two equations we shall calculate $\delta_{m}$ and $\delta_{\phi}$
for a system consisting of dark matter with negligible pressure
and quintessence dark energy with exponential potential.

\begin{figure}
\begin{center}
\psfig{file = 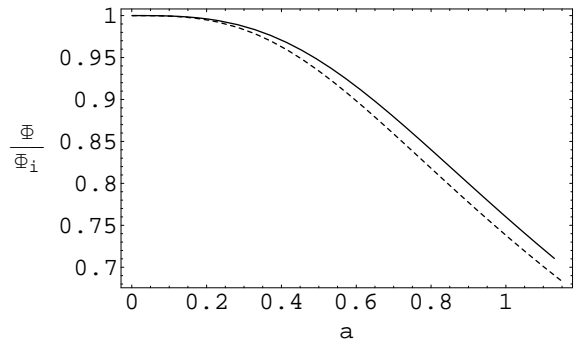, width = 3in}
\caption{This plot shows the how the gravitational potential evolve with
  time. The bold line shows how gravitational potential $\Phi_{_{N}}(a)$ evolves with
  the scale factor in the $\Lambda CDM$ model, while the dotted lines shows
  $\Phi_{_{N}}(a)$ for the dark energy model considered in this paper. Here
  the parameter $\lambda = 1$. In this plot the wave length of perturbation
  $\lambda_{p} = 10^{3}Mpc$.}
\label{plot::phi}
\end{center}
\end{figure}

\section{Numerical Solutions}\label{sec::Numerical Solutions}

For solving the background equations [Eqs.(\ref{eqn::background eqn 1}) and (\ref{eqn::background eqn 2})], we introduce the following dimensionless variables:
\begin{eqnarray}
y &=& \frac{\phi_{o}- \phi_{oi}}{M_{p}}  \label{eqn::redefined varphi}\\
s &=& \frac{a}{a_{i}}\label{eqn::redefined scale factor}\\
x &=& H_{i}(t - t_{i}) \label{eqn::redefined time}
\end{eqnarray}
where $a_{i}$, $\phi_{oi}$ and $H_{i}$ are the values of the scale factor, scalar field, and the hubble parameter at some initial time $t = t_{i}$.

In terms of these new variables, the two equations
[Eqs.(\ref{eqn::background eqn 1}) and (\ref{eqn::background eqn 2})]
describing the background cosmology becomes:
\begin{eqnarray}
\frac{s'^{2}}{s^2} - \Omega_{mi}s^{-3} - \frac{1}{3}\Big[\frac{y'{2}}{2} +
  \bar{V}\exp(-\sqrt{\lambda} y)\Big] &=& 0\label{eqn::Final eqn 1}\\
y'' + 3\frac{s'}{s}y' - \sqrt{\lambda}\bar{V}\exp(-\sqrt{\lambda} y) &=&
0\label{eqn::Final eqn 2}
\end{eqnarray}
where prime $ `` \hspace{0.1cm} ' \hspace{0.1cm} "$ corresponds to the
derivative with respect to $x$, $\bar{V} =
V_{o}H_{i}^{-2}M_{p}^{-2}\exp(-\sqrt{\lambda}M_{p}^{-1}\phi_{oi})$ and $\Omega_{mi}$
is the dimensionless matter density parameter at the initial epoch $t = t_{i}$.

Since $\Omega_{\mathrm{total}} = \Omega_{\mathrm{mi}} + \Omega_{\phi_{i}} =
1$, it follows that:
\begin{eqnarray}
\bar{V} &=& \frac{1}{2}(1 - \Omega_{mi})(1 - w_{i})\label{eqn::v bar}\\
y'_{i} &=& \sqrt{3(1 - \Omega_{mi})(1 + w_{i})}\label{eqn::yi}
\end{eqnarray}
where $w_{i}$ is the value of the equation of state parameter of the
scalar field $\phi$ at $t = t_{i}$.
Choosing $w_{i}$ to be very close to $-1$ at say red shift of $z_{i} = 1000$,
we solve the two background equations [Eqs.(\ref{eqn::background eqn 1}) and
  (\ref{eqn::background eqn 2})].

Fig.\ref{plot::omega} shows the plot of $\Omega_{m}$ and  $\Omega_{\phi}$ as a
function of scale factor. This figure gives the value of scale factor at the
matter dark energy equality $a_{eq} = 0.68$.
The corresponding redshift of matter dark energy equality is $z_{eq} = 0.46$.
The red shift at which the universe underwent a transition from
decelerated expansion phase to accelerated expansion phase turns out to be
$z_{acc} = 0.81$.
Fig.\ref{plot::w} shows the evolution of the equation of state parameter
$w_{\phi}$ as function of scale factor.
For the choice $ \lambda = 1$, this gives the value of the $w_{\phi}$ at the
present epoch to be $w_{\phi_{o}} = -0.83$.
If $ \lambda = 0.1$, then $w_{\phi_{o}} = -0.98$.
\begin{figure}[t]
\begin{center}
\psfig{file = 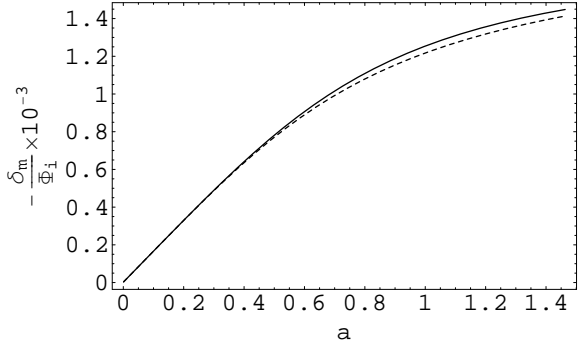, width = 3in}
\caption{This plot shows  how matter perturbation $\delta_{m}$  evolves with
  scale factor. The bold lines on this plot corresponds to $\Lambda CDM$
  model, while the dotted lines in the same plot shows how the matter
  perturbations evolves with scale factor for the dark energy model considered
  in this paper. Here the parameter $\lambda = 1$ and the wave length of
  perturbation $\lambda_{p} = 10^{3}Mpc$.}
\label{plot::deltaM}
\end{center}
\end{figure}

For numerically solving the two perturbation equations Eqs.(\ref{eqn::linerized einstein eqn 3}) and (\ref{eqn:: perturbed klein gordon eqn}), we introduce the following two dimensionless variables :
\begin{equation}
\Phi_{_{N}} = \frac{\Phi}{\Phi_{i}} \label{eqn::redefined phi}
\end{equation}

\begin{equation}
\delta y = \frac{\delta \phi}{\Phi_{i}M_{p}}
\label{eqn::redefined delta varphi}
\end{equation}
Here $\Phi_{_{N}}$ is the normalized gravitational potential with $\Phi_{i}$
being the value of the metric potential at the initial time $t = t_{i}$. In
terms of these two new variables,  the Eqs.
(\ref{eqn::linerized einstein eqn 3}) and
(\ref{eqn:: perturbed klein gordon eqn}) can be rewritten as

\begin{eqnarray}
\Phi_{_{N}}'' &+& 4\frac{s'}{s}\Phi_{_{N}}' + \Big (2\frac{s''}{s} +
\frac{s'^{2}}{s^{2}}\Big )\Phi_{_{N}} - \frac{1}{2}\Big [y'\delta y' -
  \nonumber\\
&&\Phi_{_{N}}y'^{2} +  \sqrt{\lambda} \bar{V}\exp(-\sqrt{\lambda} y)\delta
  y\Big ] = 0 \label{eqn::Final eqn 3}
\end{eqnarray}
\begin{eqnarray}
\delta y'' +  3\frac{s'}{s}\delta y' +  (\lambda \delta y -
2\Phi_{_{N}}\sqrt{\lambda})\bar{V}\exp(-\sqrt{\lambda} y) - \nonumber\\
4\Phi_{_{N}}'y' + \frac{\bar{k}^{2}\delta y}{\beta s^{2}}  = 0
\label{eqn::Final eqn 4}
\end{eqnarray}

Here $\bar{k} = k/H_{o}$, where  $H_{o}$ is the hubble parameter at the
present epoch, is the wave number scaled with respect to the
hubble radius.
In Eq. (\ref{eqn::Final eqn 4}) the constant $\beta = \frac{1}{s'^{2}_{o}}$,
where $s'_{o}$ is the value of $s'$ at the present epoch.
This can be determined numerically from the Eq. (\ref{eqn::Final eqn 1}).

We assume that the perturbation in the scalar field in the matter
dominated epoch at $z\approx 1000$ is negligibly small compared to other
perturbed quantities such as $\Phi$, $\delta_m $ etc.
The scalar field can then be treated as initially homogeneous at  $t=t_{i}$.
This corresponds to setting the initial condition $\delta y_{i} = 0$ and
$\delta y'_{i} = 0$.
The only initial condition that needs to be determined is the value of
$\Phi'_{_{N_{i}}}$ at $ t = t_{i}$.
In the matter dominated epoch ($ z \approx 1000$), the linearized Einstein's
equation (\ref{eqn::linerized einstein eqn 3}) can be analytically solved to
obtain the solution  $\dot{\Phi}(t)\propto t^{-8/3}$.
Hence $\dot{\Phi}(t)$ decays to zero in the matter dominated epoch for all
values of wave number $k$ and we can set the initial condition
$\Phi'_{_{N_{i}}}(k) = 0$.

\begin{figure}[t]
\begin{center}
\psfig{file = 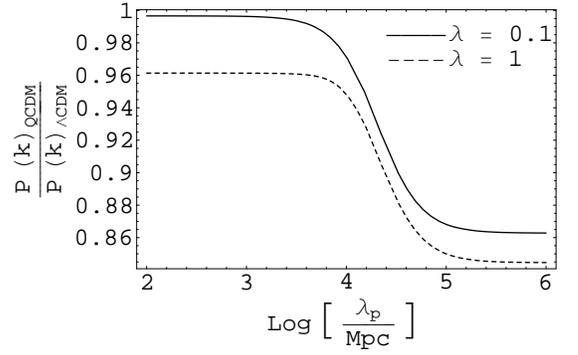, width = 3in}
\caption{The ratio of power spectrums $P(k)_{QCDM}/P(k)_{\Lambda CDM}$ as a function of length scale of perturbation $\lambda_{p}$.
If the potential parameter $\lambda = 1$, then it turns out that at the present epoch, the equation of state parameter $w_{0} = -0.83$, and if $\lambda = 0.1$ then $w_{0} = -0.98$.
In the x-axis, ``Log" refers to the logarithm to base 10.}
\label{plot :: Pk_Qcdm_by_Lcdm}
\end{center}
\end{figure}

After solving the two perturbation equations using the above initial
conditions we can now determine the dimensionless density perturbations
$\delta_{m}$ and $\delta_{\phi}$ defined in Eq.(\ref{eqn::deltaM}) and
Eq. (\ref{eqn::delta varphi}).
In terms of the dimensionless variable defined
in Eq. (\ref{eqn::redefined phi}) and Eq.
(\ref{eqn::redefined delta varphi}),
we can express $\delta_{m}$ and $\delta_{\phi}$ in the following form:
\begin{equation}
\frac{\delta_{\phi}}{\Phi_{i}} = \frac{y'\delta y' - \Phi_{_{N}}y'^{2} -
  \sqrt{\lambda}\delta y \bar{V}\exp\{- \sqrt{\lambda}y \}}{\frac{y'^{2}}{2} +
  \bar{V}\exp\{- \sqrt{\lambda}y \}}
\end{equation}
\begin{eqnarray}
\frac{\delta_{m}}{\Phi_{i}} &=& -\frac{1}{\Omega_{mi}s^{-3}}\Big
     [2\frac{s'^{2}}{s^{2}}\Phi_{_{N}} + 2\frac{s'}{s}\Phi'_{_{N}} +
       2\frac{\bar{k}^{2}}{3s^{2}\beta}\Phi_{_{N}} + \nonumber\\
&&\frac{1}{3}\Big (y'\delta y' - \Phi_{_{N}}y'^{2} - \sqrt{\lambda}\delta y
     \bar{V}\exp\{- \sqrt{\lambda}y \}\Big )\Big ]
\end{eqnarray}

In Fig. \ref{plot::phi}  we have plotted $\Phi_{_{N}}$ as a function of scale
factor for the value of the parameter  $\lambda = 1$.
The dotted line in this figure shows $\Phi_{_{N}}(a)$ for the model of dark
energy considered in this paper.
We can see that in the matter dominated epoch $\Phi_{_{N}} = 1$ and it is
constant.
Once the dark energy dominated epoch begins, the gravitational potential
starts to decay.
For comparison, in the same figure we had plotted [see bold line in
  Fig.\ref{plot::phi}] the form of $\Phi_{_{N}}(a)$ in the $\Lambda$CDM
model.
This implies that the behavior of $\Phi_{_{N}}$ in the matter
dominated era is the same in both $\Lambda$CDM model and the model of dark
energy considered in this paper.
But once the dark energy dominated phase
begins, $\Phi_{_{N}}(a)$ decays faster than the corresponding
$\Phi_{_{N}}(a)$ in the $\Lambda$CDM model.
In this figure the wavelength of perturbation was fixed to $\lambda_{p} =
10^{3}Mpc$. With $ H_{o}= 73Km/s\times Mpc^{-1}$\cite{spergel}, this value of
$\lambda_{p}$ corresponds to $\bar{k} = k/H_{o} \approx 26$.

\begin{figure}[t]
\begin{center}
\psfig{file = 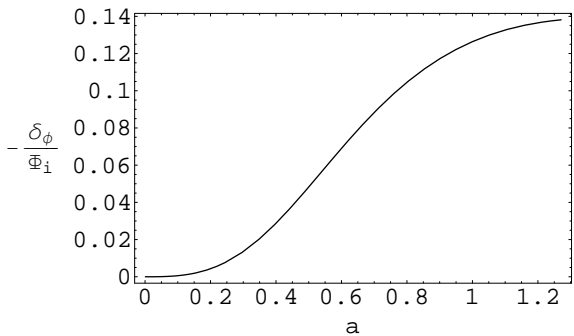, width = 3in}
\caption{This plot shows how  perturbations in dark energy  $\delta_{\phi}$
  evolve with scale factor. Here the parameter $\lambda = 1$ and the wave
  length of perturbation $\lambda_{p} = 10^{3}Mpc$.}
\label{plot::deltaDE}
\end{center}
\end{figure}
In Fig. \ref{plot::deltaM}, we have plotted the matter perturbations
$\delta_{m}$ as a function of scale factor for the dark energy potential parameter $\lambda = 1$.
In this plot the solid line corresponds to $\delta_{m}(a)$ for
$\Lambda$CDM model and the dotted
line in the same plot shows $\delta_{m}(a)$ for the dark energy model
considered in this paper.
Matter perturbation $\delta_{m}(a)$ initially grows linearly with scale
factor in the matter dominated epoch, however, once the universe undergoes
transition from the decelerated expansion phase to accelerated expansion phase,
the growth of $\delta_{m}(a)$ is suppressed.
In  fact the growth of $\delta_{m}(a)$ is suppressed more in the dark energy
model with exponential potential than the corresponding $\delta_{m}(a)$ in the
$\Lambda$CDM model.
At the present epoch, for the same set of initial conditions we find that:
\begin{equation}
\hspace{0.1cm}^{QCDM}\delta_{m}^{2}(z = 0) =  0.96\hspace{0.1cm}^{(\Lambda
  CDM)}\delta_{m}^{2}(z = 0)\label{eqn::pk}
\end{equation}
where QCDM corresponds to the quintessence + cold dark matter.
Hence, the
matter power spectrum is suppressed in QCDM model considered  in this paper than the
corresponding value in the $\Lambda$CDM model.
This  nearly $4\%$ suppression corresponds to the length scale of perturbation
$\lambda_{p} = 10^{3}Mpc$ and for the value of the parameter $\lambda = 1$.
By $4\%$ suppression, we mean that $\left[(P(k)_{\Lambda CDM} - P(k)_{QCDM})/P(k)_{\Lambda CDM}\right]\times 100  = 4.$
(Here $P(k)$ is the matter power spectrum which by definition is proportional to  $\delta_{m}^{2}$.)
In Fig.\ref{plot :: Pk_Qcdm_by_Lcdm}, we plot the ratio $P(k)_{QCDM}/P(k)_{\Lambda CDM}$ as a function of length scale of perturbation $\lambda_{p}$.
This figure implies that greater the length scale $\lambda_{p}$ and greater is the value of the parameter $\lambda$,
more is the percentage of suppression.
In fact at $\lambda_{p} = 10^{5}Mpc$, matter power spectrum is suppressed by about $15\%$ if the potential parameter $\lambda = 1$.
This implies that if $\lambda = 1$ then at $\lambda_{p} = 10^{5}Mpc$, $P(k)_{\Lambda CDM} - P(k)_{QCDM} = 0.15\times P(k)_{\Lambda CDM}$.
Since $P(k)_{\Lambda CDM}$ at $10^{3}$Mpc is at least an order of magnitude greater that the corresponding value at $10^{5}$Mpc\cite{tegmark 2003}, it follows that although the percentage of suppression is larger at larger scales, the actual difference $P(k)_{\Lambda CDM} - P(k)_{QCDM}$ is smaller at larger scales.

\begin{figure}[t]
\begin{center}
\psfig{file = 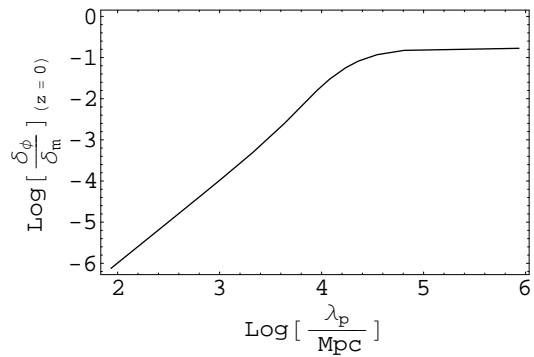, width = 3in}
\caption{This plot shows the dependence of the ratio
  $\delta_{\phi}/\delta_{m}$ at the present epoch on the wave length of
  perturbations $\lambda_{p}$. In this plot the parameter $\lambda = 1$.
  In both the axis, ``Log" refers to the logarithm to base 10.}
\label{plot::ratioL1}
\end{center}
\end{figure}
\begin{figure}[t]
\begin{center}
\psfig{file = 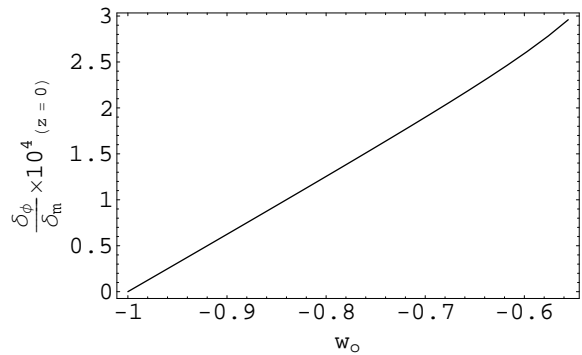, width = 3in}
\caption{This plot shows the dependence of the ratio
  $\delta_{\phi}/\delta_{m}$ at the present epoch on the equation of state
  parameter $w_{o}$. It is the value of the parameter $\lambda$ in the
  potential $V(\phi)$ which determine the present value of the equation of
  state parameter $w_{o}$. In this plot the wave length of perturbation
  $\lambda_{p} = 10^{3}Mpc$.}
\label{plot::ratio_lp3}
\end{center}
\end{figure}

\begin{figure*}
\begin{tabular}{ccc}
\begin{minipage}{3in}
\centering
\includegraphics[width=3in]{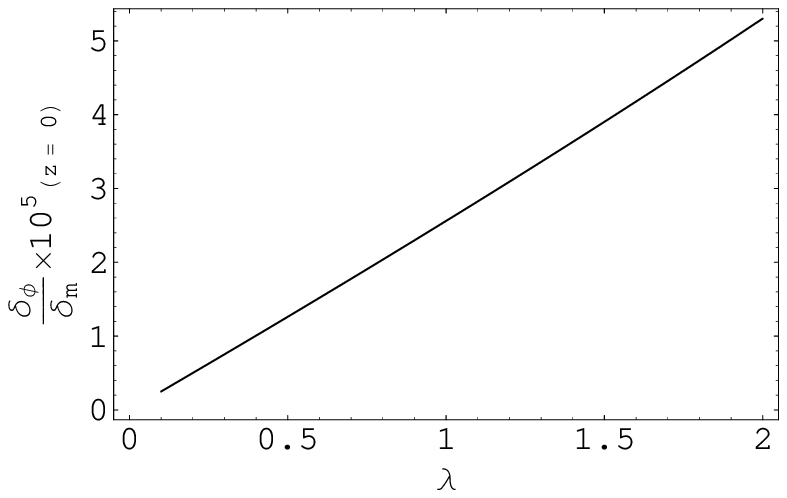}
\end{minipage}
&
\begin{minipage}{3in}
\centering
\includegraphics[width=3in]{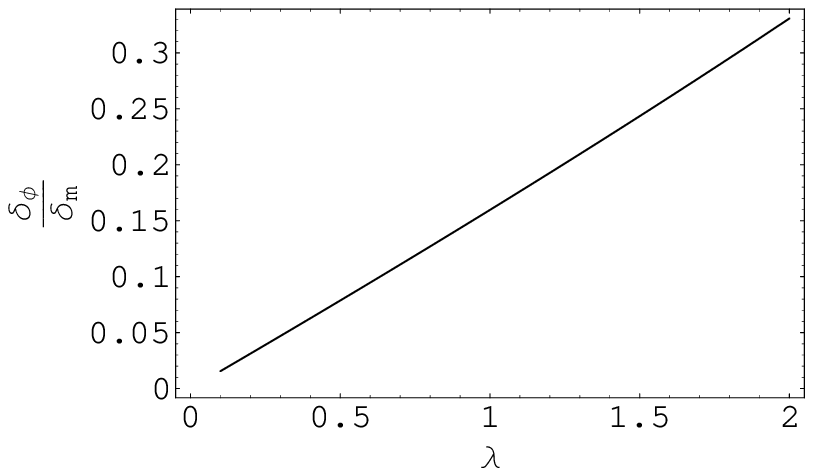}
\end{minipage}
\\
\end{tabular}
\caption{The left plot  shows the ratio of perturbation in dark energy
 to the matter perturbation at the present epoch for
length scale of perturbation of $500$Mpc as a function of potential parameter $\lambda$.
At this scale the ratio remains at almost $10^{-5}$.
The plot on the right shows the ratio at the present epoch for
length scale of perturbation of $10^{5}$ Mpc as a function of parameter
  $\lambda$.
At this length scales, for different value of the parameter $\lambda$ between $0.1$ and $2$, the ratio varies from
  0.01$-$0.33.}
\label{plot::ratio_lambda}
\end{figure*}
In Fig. \ref{plot::deltaDE}, we have plotted the perturbation in dark energy
$\delta_{\phi}$ as a function of scale factor for $\lambda = 1$.
We can see that initially, at around $z = 1000$, perturbation $\delta_{\phi}$
is almost zero.
Once the dark energy dominated epoch begins the perturbations in dark energy
grow.

\subsection{Dependence of the ratio $\delta_{\phi}/\delta_{m}$ on length
  scales}

In Figs. \ref{plot::deltaM} and \ref{plot::deltaDE} we have scaled $\delta_{m}$
and $\delta_{\phi}$ with respect to $\Phi_{i}$, the initial value of the
metric perturbation.
From Fig.\ref{plot::deltaM} and \ref{plot::deltaDE}, we find that at
the present epoch [z = 0], the ratio of dark energy perturbations to matter
perturbations at $\lambda_{p} = 10^{3}Mpc$ and  $\lambda = 1$ is given by
\begin{equation}
\Bigg(\frac{\delta_{\phi}}{\delta_{m}}\Bigg)_{(z = 0)} = 10^{-4}.
\end{equation}

Given a fixed $\lambda$, the ratio $\delta_{\phi}/\delta_{m}$ would depend on
the length scale of perturbation $\lambda_{p}$.
In Fig. \ref{plot::ratioL1}, we plot the value of this ratio
at the present epoch as a function of length scale $\lambda_{p}$.
This figure shows that at small scales  [$\lambda_{p}< 1000\mathrm{Mpc}$],
the perturbations in dark energy can be neglected in comparison with the
perturbations in matter.
This is because in  these length scales $\delta_{\phi}  \simeq
10^{-5}\delta_{m}$.
And since  $\delta_{m}$ itself is small, this value of $\delta_{\phi}$
corresponds to higher order term.
In the linear regime, for scales  $\lambda_{p}< 1000\mathrm{Mpc}$ we can
neglect the perturbations in dark energy (for this model) and we can treat
dark energy to be homogeneous.
The effect on the matter perturbation by dark energy on these  scales would be
through background $a(t)$.

On large scales [for $\lambda_{p}> 1000\mathrm{Mpc}$], the dark energy
perturbations can become comparable to $\delta_{m}$.
In Fig. \ref{plot::ratioL1}, for $\lambda_{p} = 10^{5}\mathrm{Mpc}$, we find
that $(\delta_{\phi}/\delta_{m})_{_{(z = 0)}} = 0.17.$
Even on large scales, the perturbations in dark energy can be neglected if
the equation of state parameter at the present epoch is very close to $-1$.
This matches with the fact that in a pure cosmological constant model of dark energy
with $w = -1$, its energy density is distributed homogeneously at all length scales.

In Fig.\ref{plot::ratio_lp3}, we have plotted the variation of the ratio
$\delta_{\phi}/\delta_{m}$ at the present epoch as  a function of $w_{o}$
which is the equation of state parameter at the present epoch.
Each value of $\lambda$ would result in a specific value of the equation of
state parameter $w_{o}$  as determined by the  background equations
(\ref{eqn::background eqn 1}) and (\ref{eqn::background eqn 2}) for a fixed
set of initial conditions.
This figure shows that this ratio $\delta_{\phi}/\delta_{m}\rightarrow 0$ when
$w_{0}\rightarrow -1$.
This result is true for all length scales of perturbations and it is
consistent with the argument presented in  Sec.\ref{sec::inhomogeneous DE}
that perturbation in matter implies perturbation in dark energy if $w_{de}\neq
-1$.

The two plots in  Fig. \ref{plot::ratio_lambda}, show the ratio
$\delta_{\phi}/\delta_{m}$ as function of $\lambda$ at length scale of
perturbations of $500$Mpc and $10^{5}$Mpc respectively.
Note that the figure on the left is scaled by a factor $10^{5}$.
These figures imply that  on large scales the  dependence on the parameter
$\lambda$ is stronger than on small scales.

\subsection{The role of quintessence on the matter power spectrum}

It is evident from Eq.(\ref{eqn::pk}) and Fig.\ref{plot :: Pk_Qcdm_by_Lcdm} that
matter perturbation is suppressed relative to $\Lambda$CDM model.
On length scales $\lambda_{p} < 1000\mathrm{Mpc}$, the perturbation in dark
energy is negligibly small compared to the perturbation in matter (see Fig.\ref{plot::ratioL1}).
However, even on these scales matter power spectrum is suppressed relative to
that in $\Lambda$CDM model.
This is evident from Fig.\ref{plot :: Pk_Qcdm_by_Lcdm}.
It is therefore necessary to distinguish the role of background evolution
$a(t)$ and perturbation in dark energy on the suppression of matter power
spectrum relative to $\Lambda$CDM.

In order to address this issue
we evaluate the suppression of matter power spectrum relative to $\Lambda$CDM
 if we treat dark energy as homogeneous.
In such a scenario, this suppression would be solely a consequence of different background evolution relative to $\Lambda$CDM.
Since we are considering quintessence model of dark energy, by ``\emph{homogeneous dark energy}'' we mean that we are assuming that  quintessence field is homogeneous in the longitudinal gauge.
There exists a gauge known as uniform field gauge where by definition the quintessence field is homogeneous.
However, the coordinate transformation from uniform field gauge to the longitudinal gauge would necessarily result in non zero fluctuation in the quintessence field.
Here our aim is to calculate the suppression of matter power spectrum relative to  $\Lambda CDM$ if we forcefully impose the assumption that the quintessence field is  homogeneous at all length scales and compare the same without imposing this assumption.

\begin{figure}[t]
\begin{center}
\psfig{file = 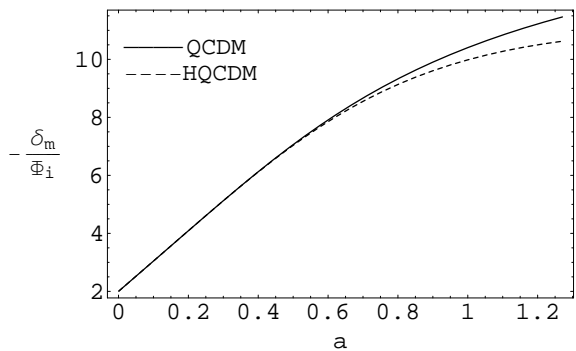, width = 3in}
\caption{In this plot the bold line (labeled as QCDM) shows how matter
  perturbation $\delta_{m}$ evolves with scale factor if we include
  perturbations in the quintessence field. For comparison the dashed line
  (labeled as HQCDM) corresponds to $\delta_{m}(a)$ assuming homogeneous
  quintessence field. In this plot $k = 5\times 10^{-4}\mathrm{Mpc}^{-1}$ and
  parameter $\lambda = 1$.}
\label{plot :: delta m qcdm hqcdm}
\end{center}
\end{figure}

If the quintessence field is homogeneous, then the evolution of the
metric perturbation $\Phi$ is determined by the following equation :
\begin{eqnarray}
\ddot{\Phi} + 4\frac{\dot{a}}{a}\dot{\Phi} + \left[\frac{\ddot{a}}{a} + \left(2 - \frac{3}{2}\Omega_{m}\left(a\right)\right)\frac{\dot{a}^{2}}{a^{2}}\right]\Phi = 0 \label{eqn :: Phi HQCDM}
\end{eqnarray}

This equation follow from Eq.(\ref{eqn::linerized einstein eqn 3}). Consequently the evolution of the matter perturbation  is determined by the following equation :
\begin{eqnarray}
\hspace{0.1cm}^{(HQCDM)}\delta_{m}^{2} &=& -\frac{2}{3H^{2}\Omega_{m}(a)}\Big\{3\frac{\dot{a}}{a}\dot{\Phi} + \frac{k^{2}\Phi}{a^{2}}+\nonumber \\
&&\left[\frac{\ddot{a}}{a} +\left(2 + \frac{3}{2}\Omega_{m}\left(a\right)\right)\frac{\dot{a}^{2}}{a^{2}}\right]\Phi\Big\}\label{eqn :: deltam HQCDM}
\end{eqnarray}
Here the superscript ``HQCDM" stands for homogeneous quintessence + cold dark matter.  The above equation [Eq.\ref{eqn :: deltam HQCDM}] follows from Eqs.(\ref{eqn::linerized einstein eqn 1}) and (\ref{eqn::linerized einstein eqn 3}).

In Fig.\ref{plot :: delta m qcdm hqcdm}, we have plotted the evolution of $\hspace{0.1cm}^{(HQCDM)}\delta_{m}(a)$ with the scale factor $a(t)$. For comparison bold line in the same figure shows $\hspace{0.1cm}^{(QCDM)}\delta_{m}(a)$, which corresponds to matter perturbations if we include quintessence fluctuations in our calculations. From this figure, it follows that although matter perturbation is suppressed relative $\Lambda$CDM, it is in fact enhanced in comparison to matter perturbation obtained by treating quintessence field as homogeneous.
Hence perturbation in dark energy actually enhances matter perturbation. This is also intuitively expected as both dark matter and dark energy clusters not anti cluster in the model considered in this paper. This is evident from  Fig.\ref{plot::deltaM} and Fig.\ref{plot::deltaDE} since both $\delta_{m}$ and $\delta_{\phi}$ have same sign. The consequent effect of this on the gravitational potential results in this enhancement of matter perturbation relative to the homogeneous dark energy case. The accelerated expansion of the of universe determined by $a(t)$ dilutes the metric perturbation $\Phi$ and consequently the overall growth of matter perturbation $\delta_{m}$ is suppressed.

\begin{figure}[t]
\begin{center}
\psfig{file = 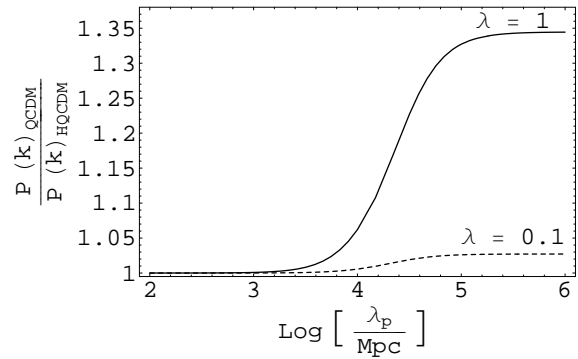, width = 3in}
\caption{This plot shows the  ratio $P(k)_{QCDM}/P(k)_{HQCDM}$ at the
  present epoch  for different length scales and for two different value of
  the parameter $\lambda$.
  In the x-axis, ``Log" refers to the logarithm to base 10.}
\label{plot :: ratio_Pk}
\end{center}
\end{figure}

In Fig. \ref{plot :: ratio_Pk}, we plot the ratio $P(k)_{QCDM}/P(k)_{HQCDM}$
at  the present epoch as function of length scale of perturbation for values
of the parameter $\lambda = 0.1$  and  $\lambda = 1$.
On length scales $\lambda_{p}< 1000\mathrm{Mpc}$, we find that $P(k)_{QCDM}
\approx P(k)_{HQCDM}$ and this is independent of the choice of parameter $\lambda$.
This implies that on these scales, including or excluding quintessence fluctuation in the perturbation equation
 does not influence the matter power spectrum significantly.
 The suppression of matter power spectrum on these scales (as shown in Fig.\ref{plot :: Pk_Qcdm_by_Lcdm})
 is therefore primarily due to different background evolution relative to that in $\Lambda$CDM model.

However, on large scales $\lambda_{p}> 1000\mathrm{Mpc}$,  $P(k)_{QCDM}$ deviates
significantly from $P(k)_{HQCDM}$  for larger value of the parameter
$\lambda$ (see Fig.\ref{plot :: ratio_Pk}).
Hence on these scales, including or excluding quintessence fluctuation in the perturbation equation
 does influence the matter power spectrum significantly.

 Fig.\ref{plot :: Pk_Qcdm_by_Lcdm} implies that $P(k)_{QCDM} < P(k)_{\Lambda CDM}$.
 However,  Fig.\ref{plot :: ratio_Pk} implies that $P(k)_{QCDM} > P(k)_{HQCDM}$.
 This means that $P(k)_{HQCDM} < P(k)_{QCDM} < P(k)_{\Lambda CDM}$.
 This implies that matter power spectrum is suppressed relative to that in  $\Lambda$CDM model
 even if treat quintessence field as homogeneous.
 This also implies that although large scale matter perturbation is suppressed in generic quintessence
dark energy model compared to that in $\Lambda$CDM,  perturbations in dark energy (in quintessence)
\emph{enhance} matter perturbation relative to the corresponding matter
perturbation obtained by treating quintessence field as homogeneous.
This enhancement is significant on large scales \textit{i.e} for $\lambda_{p}
> 1000\mathrm{Mpc}$ (see Fig. \ref{plot :: ratio_Pk}).

\begin{figure}[t]
\begin{center}
\psfig{file = 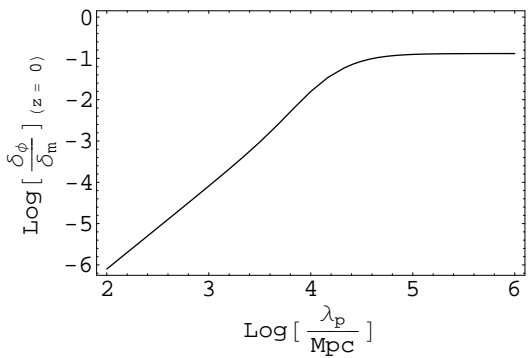, width = 3in}
\caption{In the figure we plot the ratio of dark energy perturbations to dark
  matter perturbations for $V(\phi) = 1/2 m^2 \phi^2$. The results agree with
those shown in Fig.\ref{plot::ratioL1}.
In both the axis, ``Log" refers to the logarithm to base 10.}
\label{plot :: phi2potnl}
\end{center}
\end{figure}

We compare our results with a different scalar field potential $V(\phi) =
\frac{1}{2} m^2 \phi^2$.
In Fig. \ref{plot :: phi2potnl} we plot the ratio of dark energy perturbations
to matter perturbations as a function of length scale.
The parameter $m$ is fixed to $m=0.94 H_0$ in natural units and the present day
equation of state parameter is $w = -0.87$.
The results are consistent with those shown in Fig.\ref{plot::ratioL1}.
Hence our result that quintessence dark energy can treated as homogeneous at scales $\lambda_{p}
< 1000\mathrm{Mpc}$ is a generic result.

\section{Summary and conclusions}
\label{sec::Conclusions}

In this paper,  we have investigated the perturbations in dark energy.
This is motivated by the fact that the assumption that the distribution
of dark energy (with $w \neq -1$) is homogeneous at all length scales is inconsistent with
the observational fact that dark matter is
distributed inhomogeneously.
On length scales comparable to or greater than the Hubble radius
($\lambda_{p}> 1000\mathrm{Mpc}$), the perturbation in dark energy can become
comparable to perturbation in matter if $w_{de} \neq -1$.
The model parameters we have chosen correspond to $w \approx -0.8 $ and $w
\approx -0.9$, which are within the range allowed by Supernova observations
and WMAP5 observations.
Given this range, the evolution of perturbations differs significantly.
For scales $\lambda_{p}< 1000\mathrm{Mpc}$, the perturbation in dark
energy $\delta_\phi$ can be neglected in comparison with the perturbation in
matter $\delta_{m}$ at least in the  linear regime.
We have demonstrated this using an exponential potential for the quintessence
field.
This result agrees with those presented in Ref.\cite{mota} on sub-Hubble
scales.

We have further demonstrated that quintessence dark energy results in
suppression of matter power spectrum relative  to  $\Lambda CDM$ model.
We found that at length scale of $\lambda_{p}= 1000\mathrm{Mpc}$
and for the value of the parameter $\lambda = 1$, the matter power spectrum is
suppressed by about $4\%$ compared to its value in the $\Lambda$CDM model for
the same set of initial condition.
However, at $\lambda_{p}= 10^{5}\mathrm{Mpc}$, matter power spectrum is
suppressed by about $15\%$ compared to its value in the $\Lambda$CDM model.
We have demonstrated that on scales $\lambda_{p}< 1000\mathrm{Mpc}$
this suppression is primarily due to  different background evolution relative
to $\Lambda$CDM model.
The resultant matter power spectrum is nearly invariant even if we assume that
quintessence field is homogeneous on these scales.
However, on much larger scale $\lambda_{p}> 1000\mathrm{Mpc}$, including or
excluding fluctuation in the quintessence field results in significant changes
in the  matter power spectrum.

All these results emphasize that dark energy can indeed be treated as nearly
homogeneous on scales $\lambda_{p}< 1000\mathrm{Mpc}$. However, on much larger
scales ($\lambda_{p} > 1000\mathrm{Mpc}$), if the equation of state parameter deviates from -1, then
perturbations in dark energy does influences  matter power spectrum significantly.
If a definitive detection of perturbations in dark energy is made, it will
certainly rule out the cosmological constant at least as the sole candidate of
dark energy.



\acknowledgments
We thank the anonymous referee for very useful comments and suggestions.
We also thank T. Padmanabhan, K. Subramanian, J. S. Bagla and
L. Sriramkumar for useful discussions.
SU acknowledges C.S.I.R, Govt. of India for senior research fellowships.
TRS thanks Department of Science and Technology, India for the
financial assistance.
HKJ thanks Department of Science and Technology, India for the
financial assistance through project number SR/WOS-A/PS-11/2006.
SU and TRS would like to thank  HRI, Allahabad for  hospitality where a part
of this work was carried out and  acknowledge the facilities at IUCAA
Reference Centre at Delhi University.
Computational work for this study was carried out at the cluster computing
facility at HRI, Allahabad.



\begin{thebibliography}{99}

\bibitem{obs_proof}
J.~P.~Ostriker and P.~J.~Steinhardt, Nature (London) \textbf{377}, 600 (1995);
 S.~D.~M.~White, J.~F.~Navarro, A.~E.~Evrard and C.~S.~Frenk,  Nature (London) \textbf{366}, 429 (1993);
 J.~S.~Bagla, T.~Padmanabhan, and J.~V. Narlikar, Comments on Astrophysics \textbf{18}, 275 (1996)[arXiv:astro-ph/9511102];
 G.~Efstathiou, W.~J.~Sutherland, and S.~J.~Maddox, Nature (London) \textbf{348}, 705 (1990).

\bibitem{nova_data1}
J.~L.~Tonry \textit{et al}., Astrophys.\  J.\  \textbf{594}, 1 (2003) [astro-ph/0305008];
B.~J.~Barris \textit{et~al}., Astrophys.\  J.\  \textbf{602}, 571 (2004) [astro-ph/0310843];
A.~G.~Riess \textit{et~al}., Astrophys.\ J.\  \textbf{607}, 665 (2004) [astro-ph/0402512];
P.~Astier \textit{et~al}., Astron.\ Astrophys.\ \textbf{447}, 31 (2006) [astro-ph/0510447].

\bibitem{boomerang}
A.~Melchiorri \textit{et~al}., Astrophys.\ J.\  \textbf{536}, L63 (2000) [astro-ph/9911445].


\bibitem{wmap_params}
D.~N.~Spergel \textit{et~al}.,   Astrophys.\ J.\ Suppl.\ \textbf{148}, 175 (2003) [astro-ph/0302209].

\bibitem{2003Sci...299.1532B}
S.~L.~Bridle, O.~Lahav, J.~P.~Ostriker, and P.~J.~Steinhardt, Science\ \textbf{299}, 1532 (2003).



\bibitem{spergel}
 D.~N.~Spergel \textit{et~al}., Astrophys.\ J.\ Suppl.\	 \textbf{170}, 377 (2007) [astro-ph/0603449].

\bibitem{2df}
E.~Hawkins \textit{et~al}., Mon.\ Not.\ Roy.\ Astron.\ Soc.\ \textbf{346}, 78 (2003) [astro-ph/0212375].

\bibitem{sdss}
A.~C.~Pope \textit{et~al}., Astrophys.\ J.\  \textbf{607}, 655 (2004) [astro-ph/0401249].



\bibitem{2004PhRvD..69j3501T}
M.~Tegmark \textit{et~al}., Phys.\ Rev.\ D\ \textbf{69}, 103501 (2004) [astro-ph/0310723].

\bibitem{ccprob_wein}
S. Weinberg,  Rev.\ Mod.\ Phys.\ \textbf{61}, 1 (1989).

\bibitem{review3}
T.~Padmanabhan, Phys.\ Rep.\ \textbf{380}, 235 (2003) [hep-th/0212290];
P.~J.~Peebles and B.~Ratra, Rev.\ Mod.\ Phys.\ \textbf{75}, 559 (2003) [astro-ph/0207347];
V.~Sahni and A.~Starobinsky, Int.\ J. Mod.\ Phys.\ D\ \textbf{9}, 373 (2000) [astro-ph/9904398];
J.~Ellis, Phil.\ Trans.\ Roy.\ Soc.\ Lond.\  A \textbf{361}, 2607 (2003) [astro-ph/0304183];
T.~Padmanabhan, Curr.\ Sci.\ \textbf{88}, 1057 (2005) [astro-ph/0411044].

\bibitem{quint1}
P.~J.~Steinhardt, Phil.\ Trans.\ Roy.\ Soc.\ Lond.\  A \textbf{361}, 2497 (2003);
A.~D.~Macorra and G.~Piccinelli,  Phys.\ Rev.\ D\ \textbf{61}, 123503 (2000) [hep-ph/9909459];
L.~A.~Ure{\~ n}a-L{\' o}pez and T.~Matos, Phys.\ Rev.\ D\ \textbf{62}, 081302 (2000) [astro-ph/0003364];
P.~F.~Gonz{\'a}lez-D{\'{\i}}az, Phys.\ Rev.\ D\ \textbf{62}, 023513 (2000) [astro-ph/0004125];
R.~de Ritis and A.~A.~Marino, Phys.\ Rev.\ D\ \textbf{64}, 083509 (2001) [astro-ph/0007128];
S.~Sen and T.~R.~Seshadri, Int.\ J. Mod.\ Phys.\ D\ \textbf{12}, 445 (2003) [gr-qc/0007079];
C.~Rubano and P.~Scudellaro, Gen.\ Rel.\ Grav.\ \textbf{34}, 307 (2002) [astro-ph/0103335];
S.~A.~Bludman and M.~Roos, Phys.\ Rev.\ D\ \textbf{65}, 043503 (2002) [astro-ph/0109551];
I.~Zlatev, L.~Wang and P.~J.~Steinhardt, Phys.\ Rev.\ Lett.\ \textbf{82}, 896 (1999) [astro-ph/9807002];
A.~Albrecht and C.~Skordis, Phys.\ Rev.\ Lett.\ \textbf{84}, 2076 (2000) [astro-ph/9908085].
Z.~K.~Guo, N.~Ohta and Y.~Z.~Zhang,  Phys.\ Rev.\  D {\bf 72}, 023504 (2005) [astro-ph/0505253]

\bibitem{ferreira}
P.~G.~Ferreira  and  M.~Joyce, Phys.\ Rev.\ Lett.\  \textbf{79}, 4740 (1997) [astro-ph/9707286];
P.~G.~Ferreira and M.~Joyce, Phys.\ Rev.\ D\ \textbf{58}, 023503 (1998) [astro-ph/9711102].

 \bibitem{liddle}
 A.~R.~Liddle and  R.~J.~Scherrer, Phys.\ Rev.\ D\ \textbf{59}, 023509 (1998)[astro-ph/9809272].

\bibitem{2001PhRvD..63j3510A}
C.~Armendariz-Picon, V.~Mukhanov and P.~J.~Steinhardt, Phys.\ Rev.\ D\ \textbf{63}, 103510 (2001) [astro-ph/0006373];
T.~Chiba, Phys.\ Rev.\ D\ \textbf{66}, 063514 (2002) [astro-ph/0206298];
M.~Malquarti, E.~J.~Copeland, A.~R.~Liddle and M.~Trodden, Phys.\ Rev.\ D\ \textbf{67}, 123503 (2003) [astro-ph/0302279];
L.~P.~Chimento and A.~Feinstein, Mod.\ Phys.\ Letts.\ A\ \textbf{19}, 761 (2004) [astro-ph/0305007];
R.~J.~Scherrer, Phys.\ Rev.\  Letts.\  \textbf{93}, 011301 (2004) [astro-ph/0402316].

\bibitem{tachyon1}
T.~Padmanabhan, Phys.\ Rev.\ D\ \textbf{66}, 021301 (2002) [hep-th/0204150]

\bibitem{2003PhRvD..67f3504B}
J.~S.~Bagla, H.~K.~Jassal and T.~Padmanabhan, Phys.\ Rev.\ D\ \textbf{67}, 063504 (2003) [astro-ph/0212198];
H.~K.~Jassal, Pramana\ \textbf{62}, 757 (2004) [astro-ph/0303406];
J.~M.~Aguirregabiria and R.~Lazkoz, Phys.\ Rev.\ D\ \textbf{69}, 123502 (2004) [hep-th/0402190];
A.~Sen, Phys.\ Scripta\ T\ \textbf{117}, 70 (2005) [hep-th/0312153];
V.~Gorini, A.~Kamenshchik, U.~Moschella and V.~Pasquier, Phys.\ Rev.\ D\ \textbf{69}, 123512 (2004) [hep-th/0311111];
G.~W.~Gibbons, Class.\ Quan.\ Grav.\ \textbf{20}, S321 (2003) [hep-th/0301117];
C.~Kim, H.~B.~Kim and Y.~Kim, Phys.\ Lett.\  B\ \textbf{552}, 111 (2003) [hep-th/0210101];
G.~Shiu and I.~Wasserman, Phys.\ Lett.\  B\ \textbf{541}, 6 (2002) [hep-th/0205003];
D.~Choudhury, D.~Ghoshal, D.~P.~Jatkar and S.~Panda, Phys.\ Lett.\  B\ \textbf{544}, 231 (2002) [hep-th/0204204];
A.~Frolov, L.~Kofman and A.~Starobinsky, Phys.\ Lett.\  B\ \textbf{545}, 8 (2002) [hep-th/0204187];
G.~W.~Gibbons, Phys.\ Lett.\  B\ \textbf{537} 1, (2002) [hep-th/0204008];
A.~Das, S.~Gupta, T.~Deep~Saini and S.~Kar, Phys.\ Rev.\ D\ \textbf{72}, 043528 (2005) [astro-ph/0505509];
I.~Y.~Aref'eva, astro-ph/0410443;
G. Calcagni, A. R. Liddle, Phys.\ Rev.\ D\ \textbf{74}, 043528 (2006) [astro-ph/0606003];
 E. J. Copeland, M. R. Garousi, M. Sami and S. Tsujikawa,  Phys.\ Rev.\ D\ \textbf{71}, 043003
  (2005) [hep-th/0411192].


\bibitem{2002PhLB..545...23C}
R.~R.~Caldwell, Phys.\ Letts.\  B\ \textbf{545}, 23 (2002) [astro-ph/9908168];
J.~Hao and X.~Li, Phys.\ Rev.\ D\ \textbf{68}, 043501 (2003) [hep-th/0305207];
G.~W.~Gibbons, hep-th/0302199;
V.~K.~Onemli and R.~P.~Woodard, Phys.\ Rev.\ D\ \textbf{70}, 107301 (2004) [gr-qc/0406098];
S.~Nojiri and S.~D.~Odintsov, Phys.\ Letts.\  B\ \textbf{562}, 147 (2003) [hep-th/0303117];
S.~M.~Carroll, M.~Hoffman and M.~Trodden,  Phys.\ Rev.\ D\ \textbf{68}, 023509 (2003) [astro-ph/0301273];
P.~Singh, M.~Sami and N.~Dadhich, Phys.\ Rev.\ D\ \textbf{68}, 023522 (2003) [hep-th/0305110];
P.~H.~Frampton, Mod.\ Phys.\ Letts.\ A\ \textbf{19}, 801 (2004) [hep-th/0302007];
J.~Hao and X.~Li, Phys.\ Rev.\ D\ \textbf{67}, 107303 (2003) [gr-qc/0302100];
P.~Gonz{\' a}lez-D{\'{\i}}az, Phys.\ Rev.\ D\ \textbf{68}, 021303 (2003) [astro-ph/0305559];
M.~P.~Dabrowski, T.~Stachowiak and M.~Szyd{\l}owski, Phys.\ Rev.\ D\ \textbf{68}, 103519 (2003) [hep-th/0307128];
J.~M.~Cline, S.~Jeon and G.~D.~Moore, Phys.\ Rev.\ D\ \textbf{70}, 043543 (2004) [hep-ph/0311312];
W.~Fang, H.~Q.~Lu, Z.~G.~Huang, and K.~F.~Zhang, Int.\ J.\ Mod.\ Phys.\ D\ \textbf{15}, 199 (2006) [hep-th/0409080];
S.~Nojiri and S.~D.~Odinstov, Phys.\ Rev.\ D\ \textbf{72},  023003 (2005) [hep-th/0505215];
S.~Nesseris and L.~Perivolaropoulos, Phys.\ Rev.\ D\ \textbf{70}, 123529 (2004) [astro-ph/0410309];
S.~Nojiri and S.~D.~Odinstov, Phys.\ Rev.\ D\ \textbf{70}, 103522 (2004) [hep-th/0408170];
E.~Elizalde, S.~Nojiri and S.~D.~Odinstov, Phys.\ Rev.\ D\ \textbf{70}, 043539 (2004) [hep-th/0405034];
S.~Nojiri, S.~D.~Odinstov and S.~Tsujikawa,  Phys.\ Rev.\ D\ \textbf{71}, 063004 (2005) [hep-th/0501025].
Z.~K.~Guo, N.~Ohta and Y.~Z.~Zhang, Mod.\ Phys.\ Lett.\  A {\bf 22}, 883 (2007) [astro-ph/0603109]

\bibitem{STG}
B.~Boisseau, G.~Esposito-Farese, D.~Polarski, A.~A.~Starobinsky,
Phys. Rev. Lett. \textbf{85}, 2236 (2000) [gr-qc/0001066]

\bibitem{brane1}
K.~Uzawa and J.~Soda, Mod.\ Phys.\ Letts.\ A\ \textbf{16}, 1089 (2001) [hep-th/0008197];
H.~K.~Jassal, hep-th/0312253;
C.~P.~Burgess, Int.\ J.\ Mod.\ Phys.\ D\ \textbf{12}, 1737 (2003);
%
K.~A.~Milton, Grav.\ Cosmol.\  \textbf{9}, 66 (2003) [hep-ph/0210170];
%
P.~F.~Gonz{\'a}lez-D{\'{\i}}az, Phys.\ Letts.\  B\ \textbf{481}, 353 (2000) [hep-th/0002033].

\bibitem{chaply}
A.~Y.~Kamenshchik, U.~Moschella and V.~Pasquier, Phys.\ Letts.\ B\ \textbf{511}, 265 (2001) [gr-qc/0103004];
N.~Bilic, G.~B.~Tupper and R.~D.~Viollier, Phys.\ Letts.\ B\  \textbf{535}, 17 (2002) [astro-ph/0111325];
M.~C.~Bento, O.~Bertolami and A.~A.~Sen,  Phys.\ Rev.\ D\ \textbf{66}, 043507 (2002) [gr-qc/0202064];
M.~C.~Bento, O.~Bertolami and A.~A.~Sen,  Phys.\ Rev.\ D\ \textbf{67}, 063003 (2003) [astro-ph/0210468];
M.~C.~Bento, O.~Bertolami and A.~A.~Sen, Gen.\ Rel.\ Grav.\ \textbf{35}, 2063 (2003) [gr-qc/0305086];
M.~C.~Bento, O.~Bertolami, N.~M.~C.~Santos and A.~A.~Sen, Phys.\ Rev.\ D\ \textbf{71}, 063501 (2005) [astro-ph/0412638];
O.~Bertolami, N.~M.~C.~Santos and A.~A.~Sen, Mon.\ Not.\ Roy.\ Astron.\ Soc.\ \textbf{353}, 329 (2004)[astro-ph/0402387];
A.~Dev, D.~Jain and J.~S.~Alcaniz, Phys.\ Rev.\ D\ \textbf{67}, 023515 (2003) [astro-ph/0209379].


\bibitem{water}
R.~Holman and S.~Naidu, astro-ph/0408102.

\bibitem{tp173}
T.~Padmanabhan, Class.\ Quant.\ Grav.\ \textbf{22}, 107 (2005) [hep-th/0406060];
I.~Shapiro and J.~Sola, JHEP\ \textbf{0202}, 006 (2002) [hep-th/0012227];
J.~Sola and H.~Stefancic, Phys.\ Lett.\ B\ \textbf{624}, 147 (2005) [astro-ph/0505133].

\bibitem{unified_dedm1}

T.~Padmanabhan and T.~R.~Choudhury, Phys.\ Rev.\ D\ \textbf{66}, 081301 (2002) [hep-th/0205055];
%
V.~F.~Cardone, A.~Troisi and S.~Capozziello, Phys.\ Rev.\ D\ \textbf{69}, 083517 (2004) [astro-ph/0402228];
%
P.~F.~Gonz{\'a}lez-D{\'{\i}}az, Phys.\ Letts.\  B\  \textbf{562}, 1 (2003) [astro-ph/0212414];
%
M.~A.~M.~C.~Calik, Mod.\ Phys.\ Lett.\ A\ \textbf{21},  1241 (2006) [gr-qc/0505035];
%
S.~Capozziello, S.~Nojiri and  S.~D.~Odintsov, Phys.\ Lett.\  B\ \textbf{632}, 597 (2006) [hep-th/0507182].
\bibitem{HGDE}
P.~Horava and D.~Minic, Phys.\ Rev.\ Lett.\ \textbf{85}, 1610 (2000) [hep-th/0001145];
S.~D.~Thomas, Phys.\ Rev.\ Lett.\ \textbf{89}, 081301 (2002);
M.~R.~Setare, Phys.\ Lett.\ B \textbf{653}, 116 (2007) [arXiv:0705.3517 [hep-th]];
M.~R.~Setare, Eur.\ Phys.\ J.\ C \textbf{50}, 991 (2007) [hep-th/0701085];
M.~R.~Setare, Phys.\ Lett.\ B\ \textbf{654}, 1 (2007) [0708.0118 [hep-th]].



\bibitem{2005astro.ph..5133S}
V.~K.~Onemli and R.~P.~Woodard, Class.\ Quant.\ Grav.\ \textbf{19}, 4607 (2002) [gr-qc/0204065];
T.~Padmanabhan, Phys.\ Reports\ \textbf{49}, 406 (2005) [gr-qc/0311036];
%
T.~Padmanabhan, Class.\ Quant.\ Grav.\ \textbf{19}, 5387 (2002) [gr-qc/0204019];
%
A.~A.~Andrianov, F.~Cannata and A.~Y.~Kamenshchik,Phys.\ Rev.\  D\ \textbf{72}, 043531 (2005) [gr-qc/0505087];
%
R.~Lazkoz, S.~Nesseris and  L.~Perivolaropoulos, JCAP \textbf{0511}, 010 (2005) [astro-ph/0503230];
%
M.~Szydlowski, W.~Godlowski and R.~Wojtak,Gen.\ Rel.\ Grav.\ \textbf{38}, 795 (2006)[astro-ph/0505202];

\bibitem{DEreview}
E.~J.~Copeland, M.~Sami, S.~Tsujikawa, Int.\ J.\ Mod.\ Phys.\ D\
\textbf{15}, 1753 (2006)[hep-th/0603057];
V.~Sahni, astro-ph/0403324;
T.~Padmanabhan, Phys.\ Rept.\ \textbf{380}, 235 (2003)[hep-th/0212290];
 P.~J.~E.~Peebles and B.~Ratra, Rev.\ Mod.\ Phys.\ \textbf{75}, 559 (2003)[astro-ph/0207347];
T.~Padmanabhan, arXiv:0705.2533 [gr-qc];
T.~Padmanabhan, AIP Conf.\ Proc.\  {\bf 861}, 179 (2006) [astro-ph/0603114].

\bibitem{param_fit}

H.~K.~Jassal, J.~S.~Bagla and T.~Padmanabhan, Phys.\ Rev.\ D\
\textbf{72}, 103503 (2005)[astro-ph/0506748];
 H.~K.~Jassal, J.~S.~Bagla and
T.~Padmanabhan, Mon.\ Not.\ Roy.\ Astron.\ Soc.\ \textbf{356}, L11
(2005)[astro-ph/0404378];
 H.~K.~Jassal, J.~S.~Bagla and T.~ Padmanabhan,
astro-ph/0601389;
M.~Chevallier and D~Polarski, Int.\ J.\ Mod.\ Phys.\ D\ \textbf{10}, 213 (2001) [gr-qc/0009008];
E.~V.~Linder, Phys.\ Rev.\ Lett.\ \textbf{90}, 091301 (2003) [astro-ph/0208512];
Y.~Wang, V.~Kostov, K.~Freese, J.~A.~Frieman  and P.~Gondolo, JCAP \textbf{0412}, 003 (2004) [astro-ph/0402080];
B.~A.~Bassett, P.~S.~Corasaniti and M.~Kunz, Astrophys.\ J.\ \textbf{617}, L1 (2004) [astro-ph/0407364];
S.~Lee, Phys.\ Rev.\ D\ \textbf{71}, 123528 (2005) [astro-ph/0504650];
M.~Li, Phys.\ Letts.\  B\ \textbf{603}, 1 (2004) [hep-th/0403127];
S.~Hannestad and E.~M{\" o}rtsell, JCAP \textbf{0409}, 001 (2004) [astro-ph/0407259];
L.~Perivolaropoulos,  AIP Conf.\ Proc.\  {\bf 848}, 698 (2006) [astro-ph/0601014];
Y.~Wang, arXiv:0712.0041 [astro-ph];
R.~C.~Santos, J.~V.~Cunha and J.~A.~S.~Lima, arXiv:0709.3679 [astro-ph];
L.~Samushia, G.~Chen and B.~Ratra, arXiv:0706.1963 [astro-ph];
H.~Wei and S.~N.~Zhang, Phys.\ Lett.\  B {\bf 654}, 139 (2007) [arXiv:0704.3330 [astro-ph]];
Y.~Wang and P.~Mukherjee, Phys.\ Rev.\  D {\bf 76}, 103533 (2007) [arXiv:astro-ph/0703780];
C.~Cattoen and M.~Visser, gr-qc/0703122;
Y.~Gong, A.~Wang, Q.~Wu and Y.~Z.~Zhang, JCAP {\bf 0708}, 018 (2007) [astro-ph/0703583];
C.~Clarkson, M.~Cortes and B.~A.~Bassett, JCAP {\bf 0708}, 011 (2007) [astro-ph/0702670];
S.~Nesseris and L.~Perivolaropoulos,  JCAP {\bf 0702}, 025 (2007) [astro-ph/0612653];
P.~U.~Wu and H.~W.~Yu, Phys.\ Lett.\  B {\bf 643}, 315 (2006) [astro-ph/0611507];
S.~Nojiri and S.~D.~Odintsov, J.\ Phys.\ Conf.\ Ser.\  {\bf 66}, 012005 (2007) [hep-th/0611071];
R.~J.~Colistete and R.~Giostri, astro-ph/0610916;
L.~Amendola, G.~Camargo Campos and R.~Rosenfeld, Phys.\ Rev.\  D {\bf 75}, 083506 (2007) [astro-ph/0610806];
S.~Nesseris and L.~Perivolaropoulos, JCAP {\bf 0701}, 018 (2007) [astro-ph/0610092];
V.~Sahni and A.~Starobinsky, Int.\ J.\ Mod.\ Phys.\  D {\bf 15}, 2105 (2006) [astro-ph/0610026];
L.~Samushia and B.~Ratra, Astrophys.\ J.\  {\bf 650}, L5 (2006) [astro-ph/0607301];
M.~A.~Dantas, J.~S.~Alcaniz, D.~Jain and A.~Dev,  Astron.\ Astrophys.\  {\bf 467}, 421 (2007) [astro-ph/0607060];
J.~C.~Fabris, S.~V.~B.~Goncalves, F.~Casarejos and J.~F.~Villas da Rocha, Phys.\ Lett.\  A {\bf 367}, 423 (2007) [astro-ph/0606171];
Y.~Wang and P.~Mukherjee,  Astrophys.\ J.\  {\bf 650}, 1 (2006) [astro-ph/0604051];

\bibitem{isw0}
H.~V.~Peiris and D.~N.~Spergel, Astrophys.\ J.\  {\bf 540}, 605 (2000) [astro-ph/0001393].
\bibitem{amendola1}L. Amendola, M. Kunz and D. Sapone, arXiv:0704.2421.



\bibitem{mota}
D.~F.~Mota, D.~J.~Shaw and J.~Silk, arXiv:0709.2227 [astro-ph]


\bibitem{weller_lewis}
J.~Weller and A.~M.~Lewis, Mon.\ Not.\ Roy.\ Astron.\ Soc.\ \textbf{346},  987
  (2003)[arXiv:astro-ph/0307104];

\bibitem{bean_dore}
R.~Bean and  O.~Dore,
  Phys.\ Rev.\ D \textbf{69}, 083503 (2004)[astro-ph/0307100];


\bibitem{depert}
N.~Bartolo, P.~S.~Corasaniti, A.~R.~Liddle and M.~Malquarti,
  Phys.\ Rev.\ D\ \textbf{70}, 043532 (2004)[arXiv:astro-ph/0311503];
W.~Hu, Phys.\ Rev.\ D \textbf{71}, 047301 (2005)[arXiv:astro-ph/0410680];
C.~Gordon and W.~Hu, Phys.\ Rev.\ D \textbf{70},
  083003 (2004)[arXiv:astro-ph/0406496];
C.~Gordon, Nucl.\ Phys.\ Proc.\ Suppl.\ \textbf{148},
  51 (2005)[arXiv:astro-ph/0503680];
C.~Gordon and  D.~Wands, Phys.\ Rev.\ D\
  \textbf{71}, 123505 (2005)[arXiv:astro-ph/0504132];
L.~R.~Abramo and  F.~Finelli, Phys.\ Rev.\ D \textbf{64},
  083513 (2001)[astro-ph/0101014];

\bibitem{chpgas_pert}
 J.~C.~Fabris, S.~V.~B.~Goncalves and P.~E.~de Souza, Gen.\ Rel.\ Grav.\  {\bf 34}, 53 (2002)
  [arXiv:gr-qc/0103083];
W.~Zimdahl and J.~C.~Fabris, Class.\ Quant.\ Grav.\  {\bf 22}, 4311 (2005)
  [arXiv:gr-qc/0504088];
S.~Silva e Costa, M.~Ujevic and A.~Ferreira dos Santos, arXiv:gr-qc/0703140;
V.~Gorini, A.~Y.~Kamenshchik, U.~Moschella, O.~F.~Piattella and A.~A.~Starobinsky, arXiv:0711.4242 [astro-ph];




\bibitem{sph_coll}
S.~Dutta and I.~Maor, Phys.\ Rev.\  D {\bf 75}, 063507 (2007) [arXiv:gr-qc/0612027];
I.~Maor and O.~Lahav, JCAP {\bf 0507} (2005) 003
  [arXiv:astro-ph/0505308];
P.~Wang, Astrophys.\ J.\  {\bf 640}, 18 (2006)
  [arXiv:astro-ph/0507195];
 D.~F.~Mota and C.~van de Bruck, Astron.\ Astrophys.\  {\bf 421}, 71 (2004)
  [arXiv:astro-ph/0401504];
N.~J.~Nunes and D.~F.~Mota, Mon.\ Not.\ Roy.\ Astron.\ Soc.\  {\bf 368}, 751 (2006)
  [arXiv:astro-ph/0409481];
C.~Horellou and J.~Berge, Mon.\ Not.\ Roy.\ Astron.\ Soc.\  {\bf 360}, 1393 (2005)
  [arXiv:astro-ph/0504465];
 L.~R.~Abramo, R.~C.~Batista, L.~Liberato and R.~Rosenfeld, JCAP {\bf 0711}, 012 (2007)
  [arXiv:0707.2882 [astro-ph]].


\bibitem{ratra88}
B.~Ratra and P.~J.~E.~Peebles, Phys.Rev. D \textbf{37}, 3406
  (1988).

\bibitem{burd}
A.~B.~Burd and J.~D.~Barrow, Nucl. Phys. B \textbf{308}, 929
  (1988).

\bibitem{copeland}
E.~J.~Copeland, A.~R.~Liddle and  D.~Wands, Phys. Rev. D
  \textbf{57}  4686 (1998) [arXiv:gr-qc/9711068].

\bibitem{exp_pot}
J.J. Halliwell, Phys. Lett. B \textbf{185}, 341 (1987);
F. Lucchin and S. Matarrese, Phys. Rev. D \textbf{32}, 1316 (1985);
C. Wetterich, Astron. Astrophys. \textbf{301}, 321 (1995)
[arXiv:hep-th/9408025].

\bibitem{bardeen PRD 1980}
J. Bardeen,  Phys. Rev. D  \textbf{22}, 1882 (1980)

\bibitem{Kodama}
H. Kodama and M. Sasaki, Prog. Theor. Phys. Suppl., \textbf{78}, 1 (1984).

\bibitem{mukhanov 1992}
V. F. Mukhanov, H. A. Feldman and R. H. Brandenberger,  Phys. Rep. \textbf{215}, 203 (1992).

\bibitem{tp_rev}
T.~Padmanabhan, astro-ph/0602117.
\bibitem{tegmark 2003}
M.~Tegmark \textit{et~al}., Astrophys.J. \textbf{606}, 702 (2004) [arXiv:astro-ph/0310725v2].

\end{thebibliography}
\end{document}